\documentclass{appolb}
\usepackage{graphicx}
\usepackage[utf8]{inputenc}
 
\begin{document}
\title{A new model for soft interactions in Herwig%
\thanks{Presented at EPIPHANY XXIII}%
}
\author{Stefan Gieseke, Patrick Kirchgaesser\footnote{Speaker}, Frasher Loshaj
\address{Institute for Theoretical Physics, Wolfgang-Gaede Strasse 1, 76131 Karlsruhe, Germany.}
\\
\address{Karlsruhe Institute of Technology}
}
\maketitle
\begin{abstract}
We present a new model for soft interactions in the Monte Carlo event-generator Herwig. The soft diffractive final states are modeled on the basis of the cluster hadronization model and interactions between soft particles are modeled as multiple particle production with multiperipheral kinematics.
We further present much improved results of mininum-bias measurements at different energies.
\end{abstract}
  
\section{Introduction}
A typical proton-proton collision consists of many different aspects. Besides the hard interaction, additional activity, so-called multiple particle interactions have to be considered in order to be able to simulate and predict minimum-bias measurements and data coming from the underlying-event.
While there has been tremendous efforts in describing the hard interaction with perturbation theory, the attempts to find a theoretical description of the soft regime, where interactions happen at energy scales well below perturbation theory is valid, have been stagnating for quite some time and Monte-Carlo event generators often limit themselves to a mere modeling of final states.
In order to simulate physics in the soft regime one heavily relies on models which should, at least to some extend, be inspired by theoretical and phenomenological considerations.
This talk deals with the implementation of such a model into the Monte-Carlo event generator Herwig\,\cite{Bahr:2008pv,Bellm:2015jjp}.
With the entering of the LHC into the high-luminosity era, and the rise of increasingly accurate data, a thorough understanding of all effects who contribute to the total cross section is necessary in order to make reasonable predictions.


\subsection{The MPI model in Herwig}

To model the underlying event, the so-called JIMMY package which was an add-on for the FORTRAN version of Herwig\,\cite{Butterworth:1996zw} was used. A similar model for hard multiple parton interactions was implemented into the newer C++ version of Herwig\,\cite{Bahr:2008wk,Bahr:2008dy}. Besides additional hard interactions, soft interactions were introduced within the so-called hot-spot model\,\cite{Borozan:2002fk,Bahr:2009ek}.
There are two main parameters in the model, the minimum transverse momentum $p_{\perp}^{\rm{min}}$ at which hard hard multiple parton interactions appear and the inverse proton radius $\mu^2$ which governs the transverse spatial distribution of partons within the hadron.
The additional soft interactions are modeled with the same functional form of the spatial distribution but another proton radius $\mu^2_{\rm{soft}}$ is used to take the different, more broader, distribution of soft particles within the hadron into account. The transverse momentum of these soft particles is then modeled with a Gaussian below the cut-off scale for the hard interactions $p_{\perp}^{\rm{min}}$ ,
\begin{equation}
\frac{\rm{d}\sigma}{\rm{d}p_{\perp}} = \left( \frac{\rm{d}\sigma_{\rm{hard}}}{\rm{d}p_{\perp}}\right)_{p_{\perp} = p_{\perp}^{\rm{min}}}\left( \frac{p_{\perp}}{p_{\perp}^{\rm{min}}}\right) e^{-\beta(p_{\perp}^2 - p_{\perp}^{\rm{min}2})}.
\end{equation}
The spectrum is chosen in such a way that it is continuous at $p_{\perp}^{\rm{min}}$. The two parameters of the soft model $\mu^2_{\rm{soft}}$ and $\beta$ are fixed in order to describe the total cross section $\sigma_{\rm{tot}} = \sigma_{\rm{hard}}^{\rm{inc}} + \sigma_{\rm{soft}}^{\rm{inc}}$ which is given by the Donnachie-Landshoff parametrization\,\cite{Donnachie:1992ny} and the elastic slope parameter.

\begin{figure}
\centering
	\includegraphics[scale=0.5]{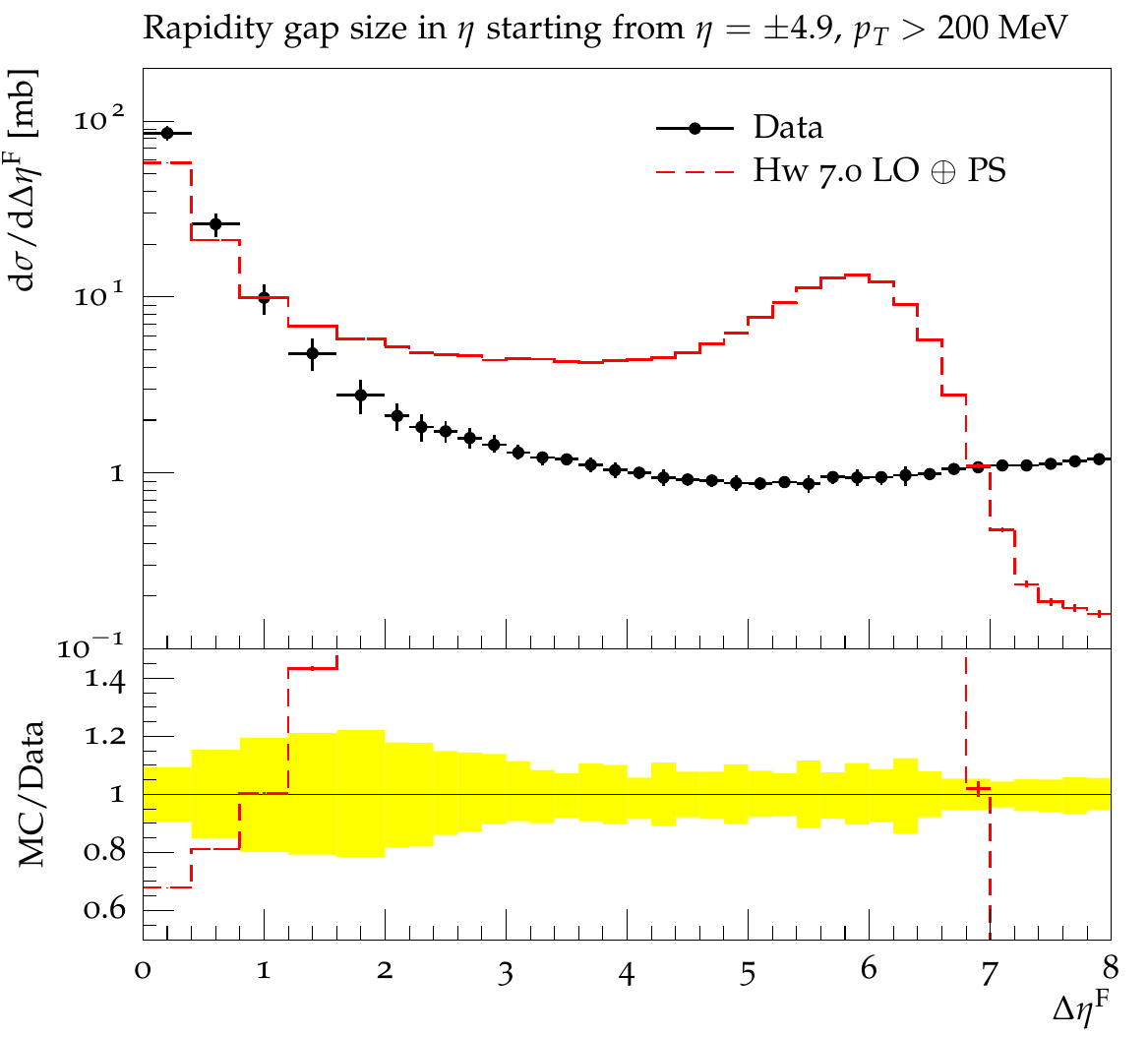}
	\caption{ATLAS rapidity gap measurement at $\sqrt{s} = 7\,\rm{TeV}$ with $p_{\perp}>200\,\rm{MeV}$ in the range $| \eta |\leq 4.9$ \cite{Aad:2012pw}.}
	\label{bump}
\end{figure}

This model was able to give a good description of the underlying-event in the presence of at least one hard scattering event. In order to model minimum-bias data a dummy process where two quarks with zero transverse momentum are extracted from the protons is introduced and the secondary hard and soft scatters are then generated from the proton remnant. The number of these additional soft scatterings is calculated in the eikonal model and simulated as an elastic scattering among gluons with $p_{\perp}<p_{\perp}^{\rm{min}}$.
This model gave good results when the hard contribution dominated or when diffractive events where suppressed by cutting on the number of final state charged particles\,\cite{Gieseke:2012ft}. Arbitrary colour connections between the gluons and the remnants produced events with large gaps in rapidity why Herwig was found to overemphasize those events\,\cite{Aad:2012pw,Gieseke:2016pbi} as can be seen in Fig.\,\ref{bump}.
This triggered the development of a model for diffracive final states and a new model for soft interactions\,\cite{Gieseke:2016pbi} which will be explained in the following chapters.





\section{Model for diffractive final states}

\begin{figure}
\centering
	\includegraphics[scale=0.6]{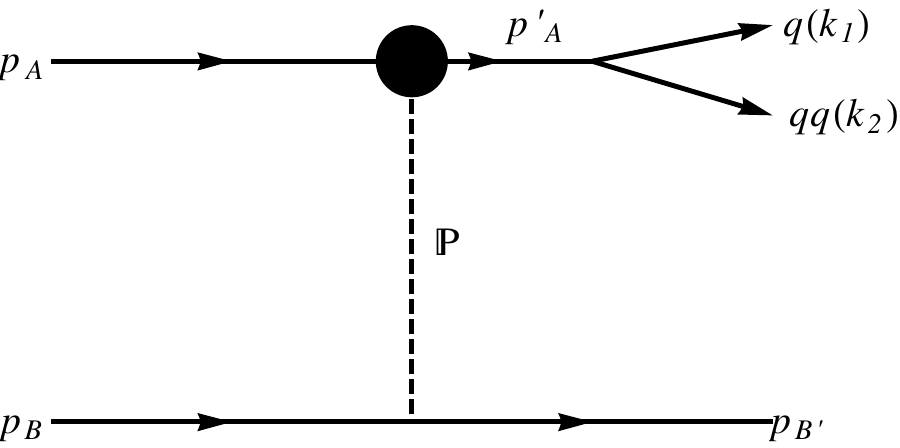}\\
	\includegraphics[scale=0.6]{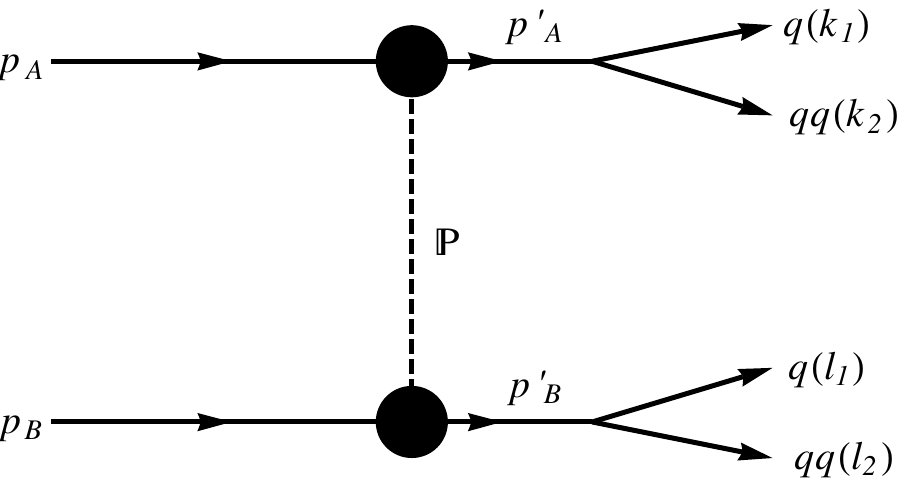}
	\caption{Diffractive dissociation for single (top) and double (bottom) diffraction.}
	\label{matrixelements}
\end{figure}

We generate single- and double diffractive events according to the differential cross sections which can be described by Regge theory and the genralized optical theorem.
The process for single diffraction can be depicted by $A+B \rightarrow X+B$ where $A$ and $B$ are the incoming hadrons and $X$ is some hadronic final state in the limit $s \gg M^2 \gg |t|$. $s$ is the center of mass energy and $M$ is the invariant mass of the diffractive final state $X$. The same holds true for the double diffractive process $A + B \rightarrow X_A + X_B$.
The two processes are depicted in Fig.\,\ref{matrixelements}.
The cross sections for hard and soft interactions only sum up to a certain fraction of the total cross section. The contribution from diffractive events is assumed to be roughly about a rate of 20 -25 $\%$ of the total event rate which has to be taken into account when integrating the diffractive model into the existing model for multiple particle interactions in Herwig.
After sampling the invariant mass and the scattering angle, the outgoing momenta of the diffractive states are constructed. The dissociated proton is then decayed further into a quark-diquark pair moving collinear to the dissociated proton. A cluster is formed out of this quark-diquark pair and then handled further by the hadronization model. A small fraction of the diffractive events for very low diffractive masses is modeled with the $\Delta$ baryon as a final state instead of a quark-diquark pair. For single diffraction $p + p \rightarrow \Delta + p$ and for double diffraction $p + p \rightarrow \Delta + \Delta$. $\Delta$ is then handled by the decay handler.
 
\section{New model for soft interactions}

\begin{figure}
\centering
	\includegraphics[scale=1.0]{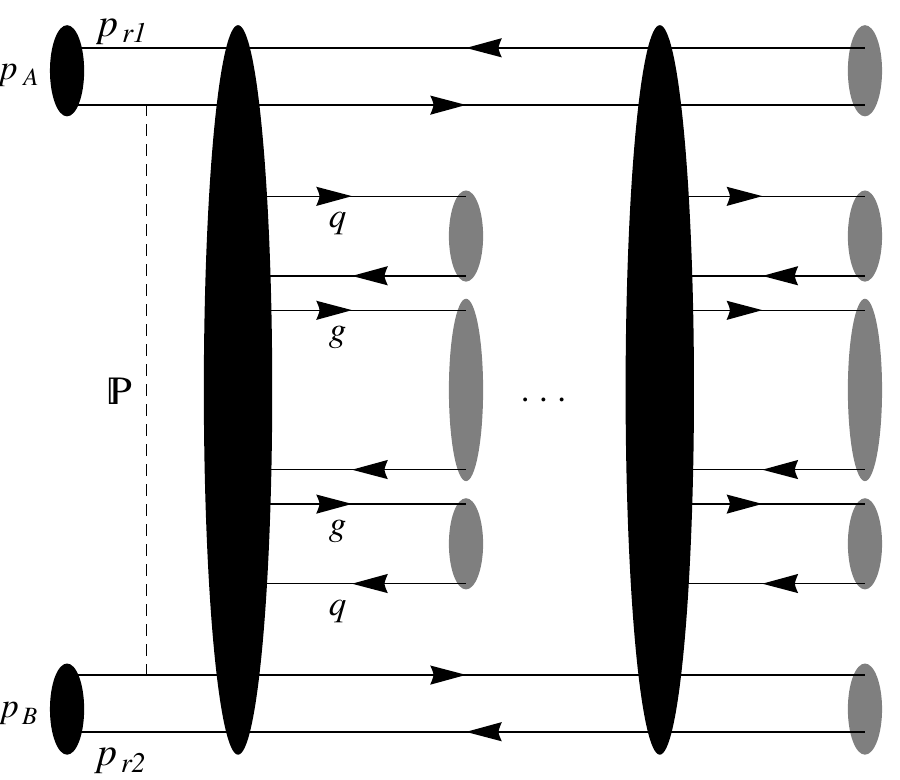}
	\caption{Cluster formation in the multiperipheral final state with mutltiple interactions.}
	\label{ladder}
\end{figure}

The new model addresses some of the short-comings of the old model for soft interactions and improves the description of many minimum-bias observables significantly.
The kinematics of the soft particles is constructed according to a multiperipheral particle production which was introduced in\,\cite{Baker:1976cv}.
In adapting this model to our pre-existing model for multiple particle interactions we view the number of soft interactions as Pomeron exchanges which leads to production of n-particle ladders if cutted.
We note that the final particles that appear in the ladder originate from the proton remnants and are modeled as sea quarks and gluons. 
The multiperipheral ladder is illustrated in Fig.\,\ref{ladder} where the dashed line represents a Pomeron exchange between the two quarks. The total energy available to produce the particles within the ladder is given by the total energy of the proton remnants, which are denoted $p_{r1}$ and $p_{r2}$. The number $N$ of the particles in the ladder is sampled from a Poissonian distribution with mean at
\begin{equation}
	\langle N \rangle = n_{\rm{ladder}} \ln \frac{(p_{r1} + p_{r2})^2}{m_{\rm{rem}}^2},
\end{equation}
where $m_{\rm{rem}}$ is the constituent mass of the proton remnant and $n_{\rm{ladder}}$ is a constant which is tuned to data.
The momentum fraction given to each particle is calculated in such a way that the particles are distributed equally in rapidity space. For more details on how the algorithm works, see Ref.\,\cite{Gieseke:2016pbi}.
In the cluster model colour connected partons form a cluster which are illustrated as gray blobs in Fig.\,\ref{ladder}. The initial quark that is extracted from the proton remains colour connected to the remnant and forms a cluster. The sea quark, denoted by q is colour connected to the first gluon in the ladder denoted by g. The subsequent generated gluons are colour connected to each other in order to form clusters that are also equally spaced in rapidity and do not range over a large rapidity interval individually.
With this algorithm it is possible to guarantee an exponential fall off of the amplitude for large values of the rapidity gap $\Delta \eta$. It also produces a roughly flat distribution in rapidity space of the produced clusters and the subsequently generated particles.
The probability for having $k$ such soft interactions (where each one produces a multiperipheral ladder of particles) is computed by the preexisting implementation of the MPI model in Herwig. 

\section{Results}
\subsection{Rapidity gap analysis} 
\begin{figure}
\centering
	\includegraphics[scale=0.5]{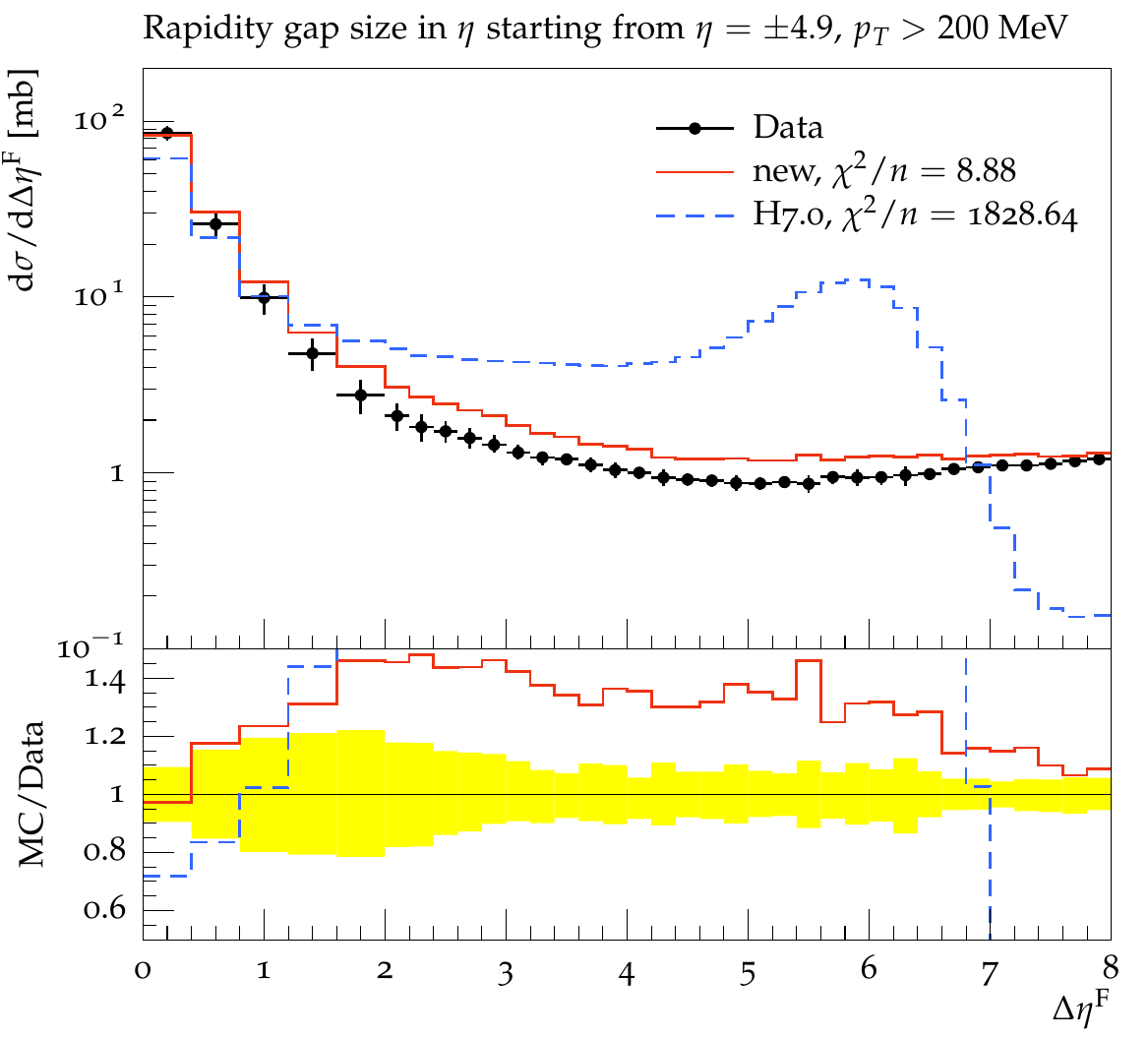}
	\includegraphics[scale=0.5]{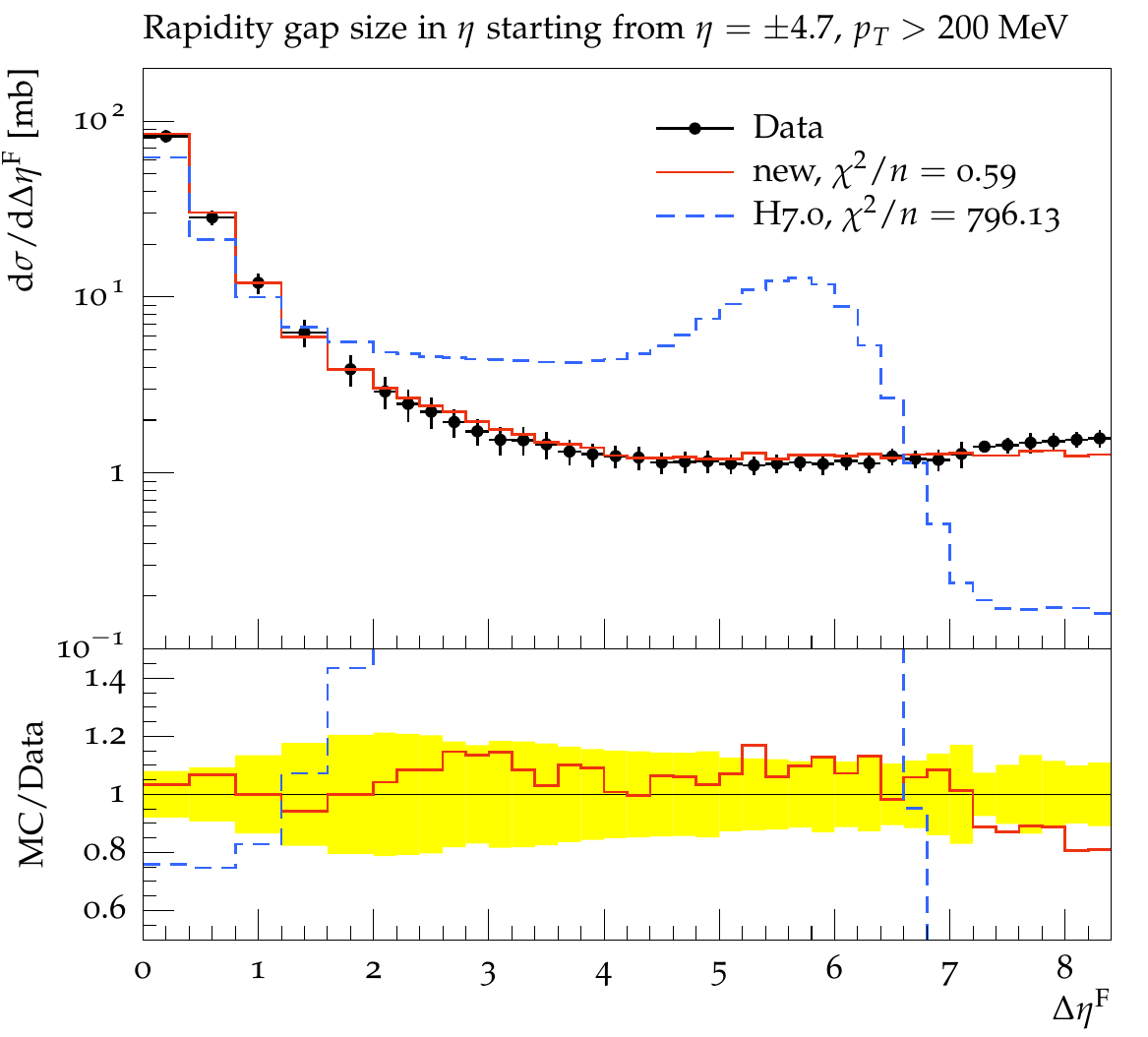}
	\caption{Comparison of the new model for soft interactions and diffraction with the old model from Herwig 7 to rapidity gap measurements at $\sqrt{s} = 7\, \rm{TeV}$ from ATLAS\,\cite{Aad:2012pw} (left) and CMS\,\cite{Khachatryan:2015gka} (right).}
	\label{rapiditygap}
\end{figure}

In Ref.\,\cite{Aad:2012pw,Khachatryan:2015gka} the differential cross section w.r.t the rapidity gap size in forward direction $\Delta \eta^F$ is measured. $\Delta \eta^F$ is defined as the larger of two pseudorapidity regions in which no particles are produced. The range of $\Delta \eta^F$ is restricted by the geometry of the detector and therefore different for ATLAS and CMS. All particles with $p_{\perp} > p_{\perp}^{\rm{cut}}$ are analyzed where $p_{\perp}^{\rm{cut}}$ varies from $200\,\rm{MeV}$ to $800\,\rm{MeV}$. This observable can be decomposed into a non-diffractive and a diffractive part of the differential cross section where small gaps are mainly dominated by non-diffractive contributions and larger gaps by single- and double diffractive events.
For non-diffractive processes the differential cross section decreases exponentially w.r.t the rapidity gap $\rm{d}\sigma/\rm{d}\Delta \eta^F \sim \exp(-a\Delta \eta^F)$ where $a$ is some constant. Diffractive events which result from pomeron exchange give rise to a constant differential cross section w.r.t $\Delta \eta^F$. By combining the models for soft interactions and diffraction Herwig  is able to describe the measurement of the rapidity gap by ATLAS and CMS as shown for two examples in Fig.\,\ref{rapiditygap}. Despite similar cuts the simulation overestimates the the data provided by ATLAS.

\subsection{Minimum-bias data}

\begin{figure}
\centering
	\includegraphics[scale=0.5]{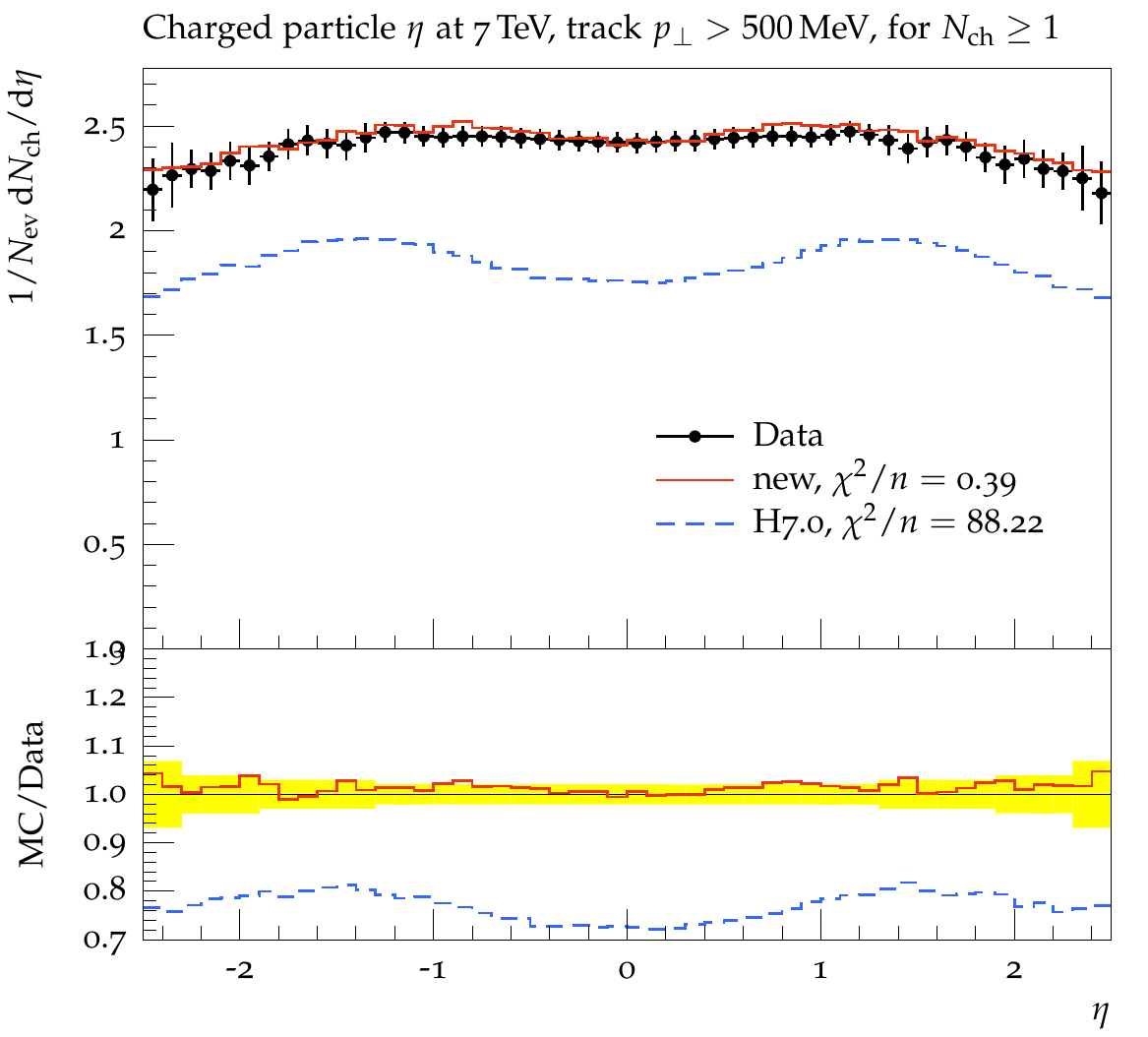}
	\includegraphics[scale=0.5]{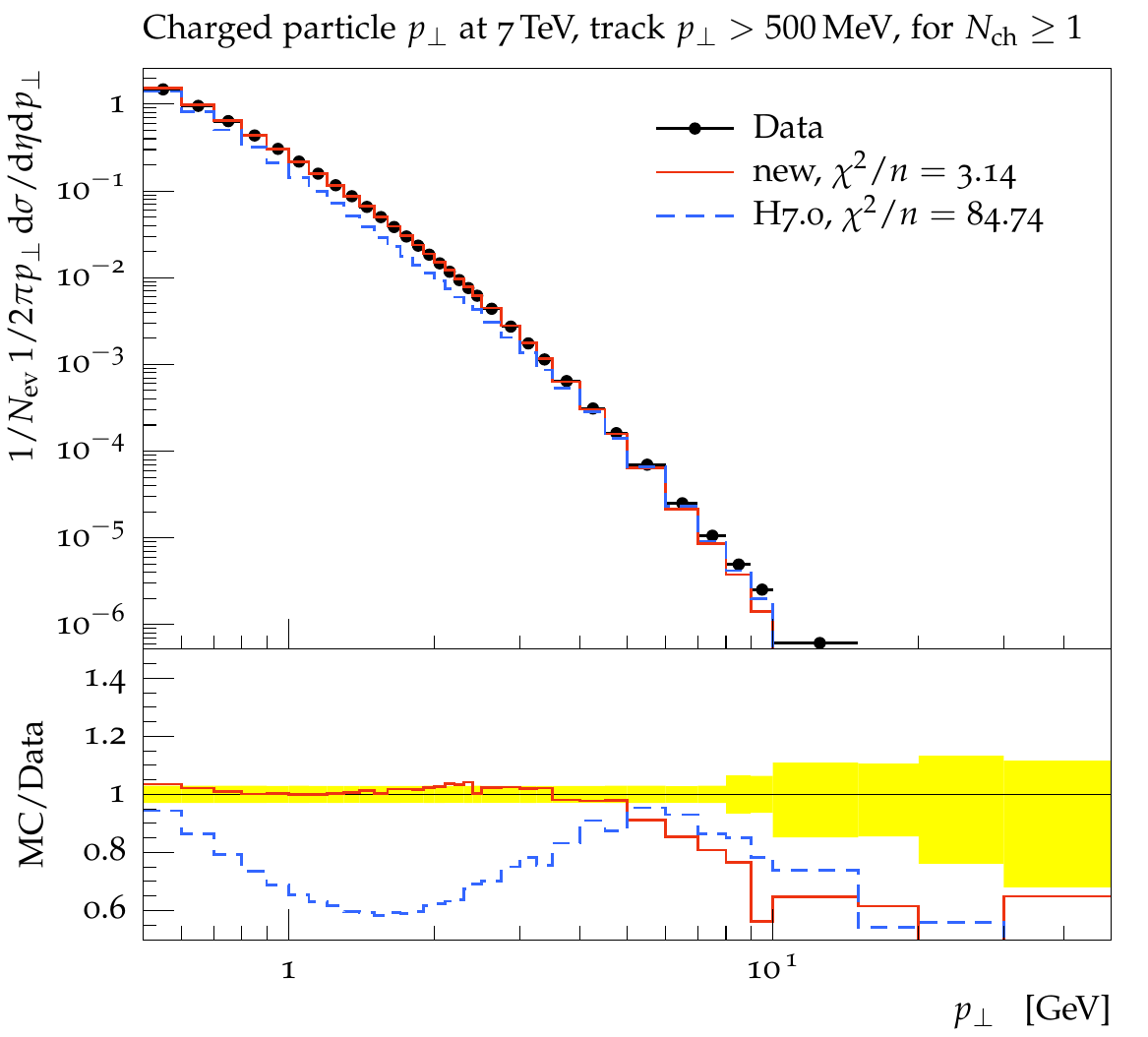}
	\caption{Comparison of the default tune from Herwig 7.0 with the new model for soft interactions to minimum-bias data from ATLAS\,\cite{:2010ir}.}
	\label{minbias}
\end{figure}

The new models for soft interactions and diffraction were tuned to minimum-bias data from ATLAS \cite{:2010ir,Aad:2016mok} at $\sqrt{s} = 900 \, \rm{MeV}$, $\sqrt{s} = 7\, \rm{TeV}$ and $\sqrt{s} = 13\, \rm{TeV}$. The results for the Monte-Carlo runs with the tuned parameters for $7\,\rm{TeV}$ are shown in Fig.\,\ref{minbias}. Here we show the most inclusive $\eta$ distribution and the distribution for the charged particle $p_{\perp}$ in the region we deem sensitive to the effects of soft particle production. We note that the overall description improves significantly compared to H7.0. In combination with the two new models Herwig is for the first time able to describe almost all aspects of minimum-bias data.

\subsection{Underlying-event}

\begin{figure}
\centering
	\includegraphics[scale=0.5]{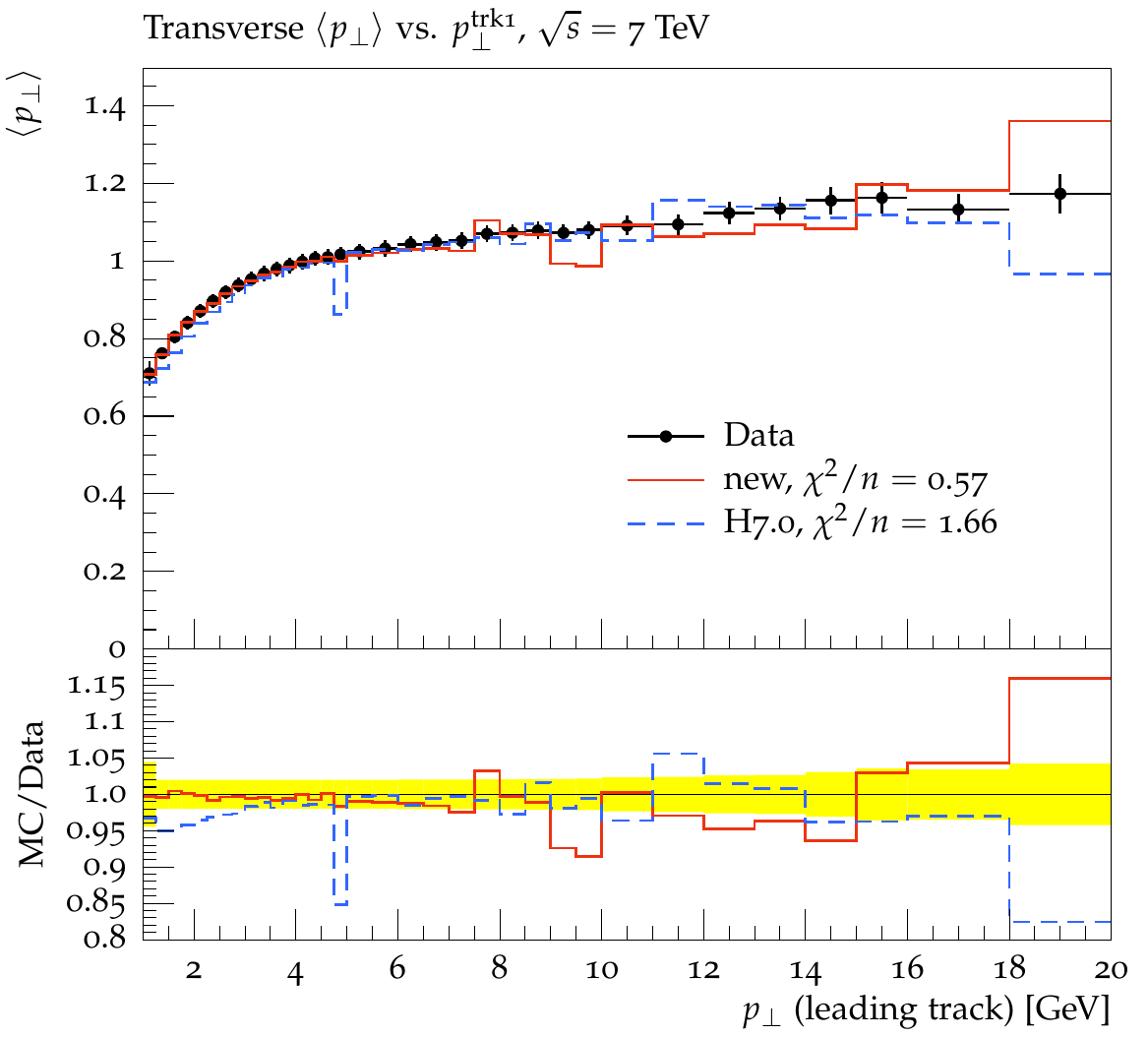}
	\includegraphics[scale=0.5]{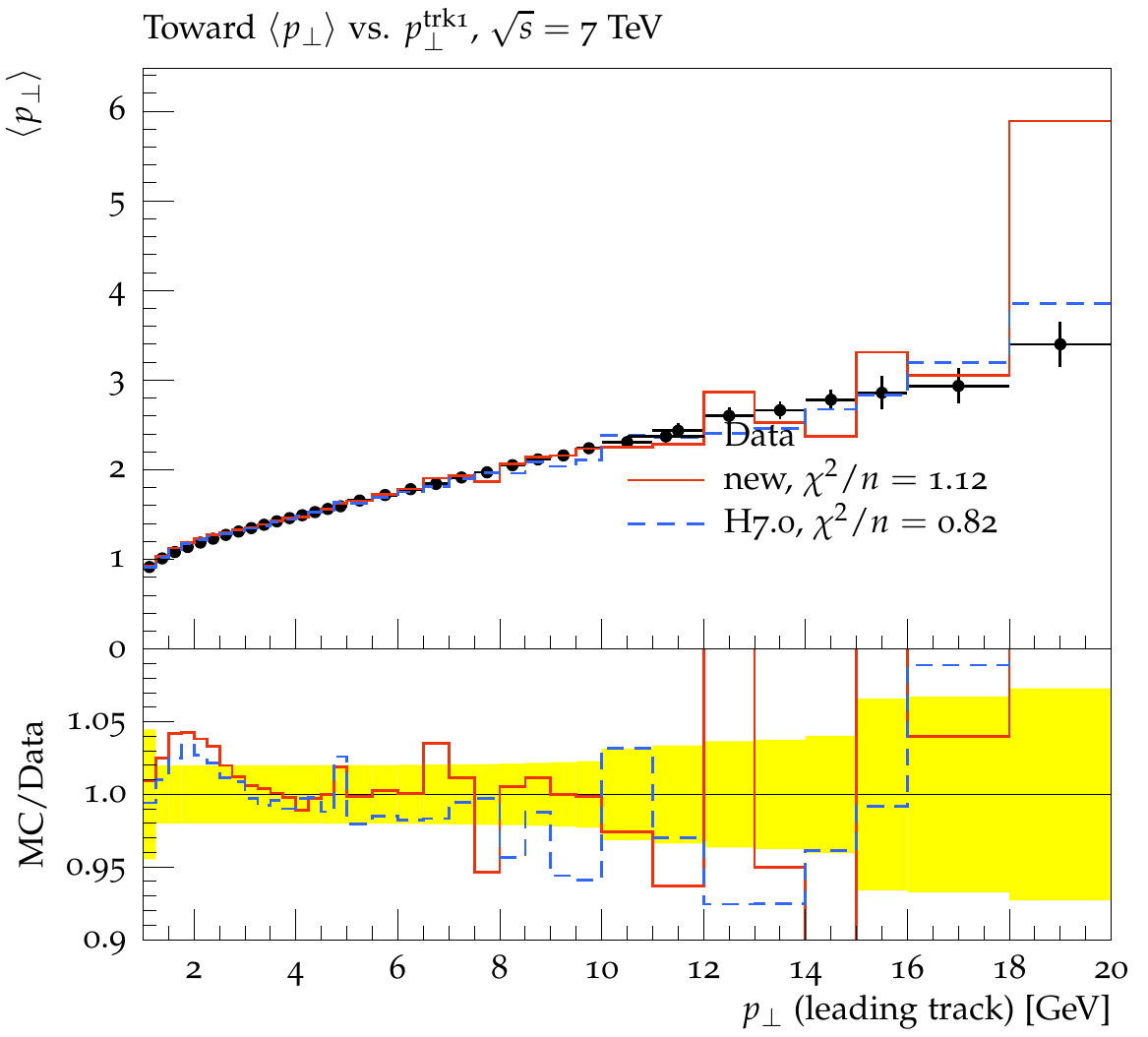}
	\caption{Average transverse momentum $\langle p_{\perp} \rangle$ as a function of $p_{\perp}^{\rm{lead}}$ for the transverse and the forward region cite .}
	\label{underlyingevent}
\end{figure}

The next important question is to whether the new models affect our previously  satisfying description of the underlying event data \cite{Gieseke:2012ft}.
The underlying event is characterized as "everything except the hard scattering process" and consists of contributions from initial- and final-state radiation, as well as hard and soft multiple particle interactions. Underlying-event measurements are usually separated into different regions which are defined relative to a leading object which is in this case the hardest charged track. Three regions are defined in terms of the azimuthal angle w.r.t the leading track. The forward region and the away region which are dominated by activity of the triggered hard scattering process and the transverse region, which on the other hand contains little contribution from the hard process and is sensitive to activity from the underlying event. 
In Fig.\,\ref{underlyingevent} we show the average transverse momentum $\langle p_{\perp} \rangle$ w.r.t the $p_{\perp}$ of the leading track for the transverse and the towards region. We see that especially for low $p_{\perp}$ the data is described reasonably well.

\subsection{Extrapolation to 13 TeV}
\begin{figure}
\centering
	\includegraphics[scale=0.5]{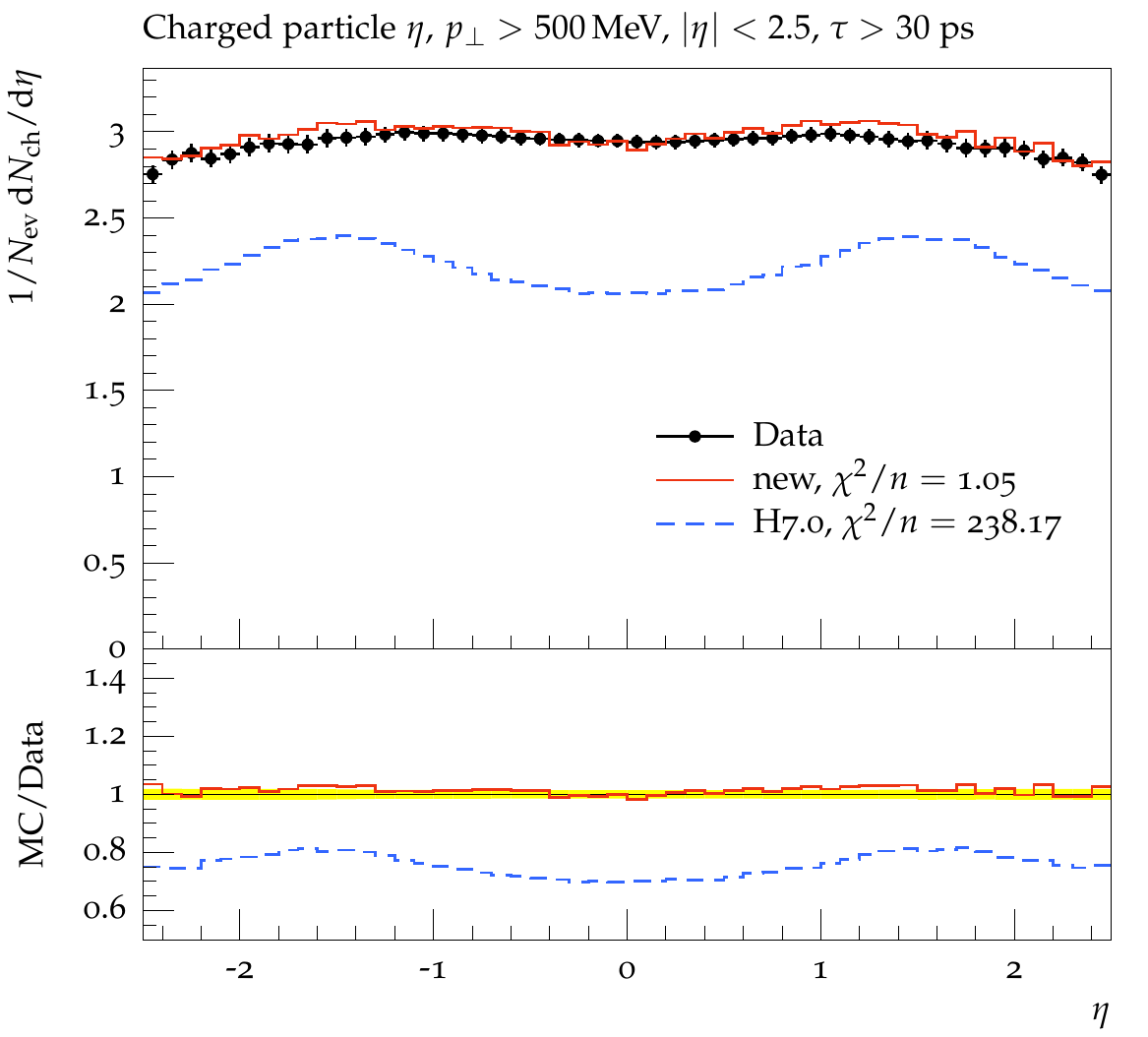}
	\includegraphics[scale=0.5]{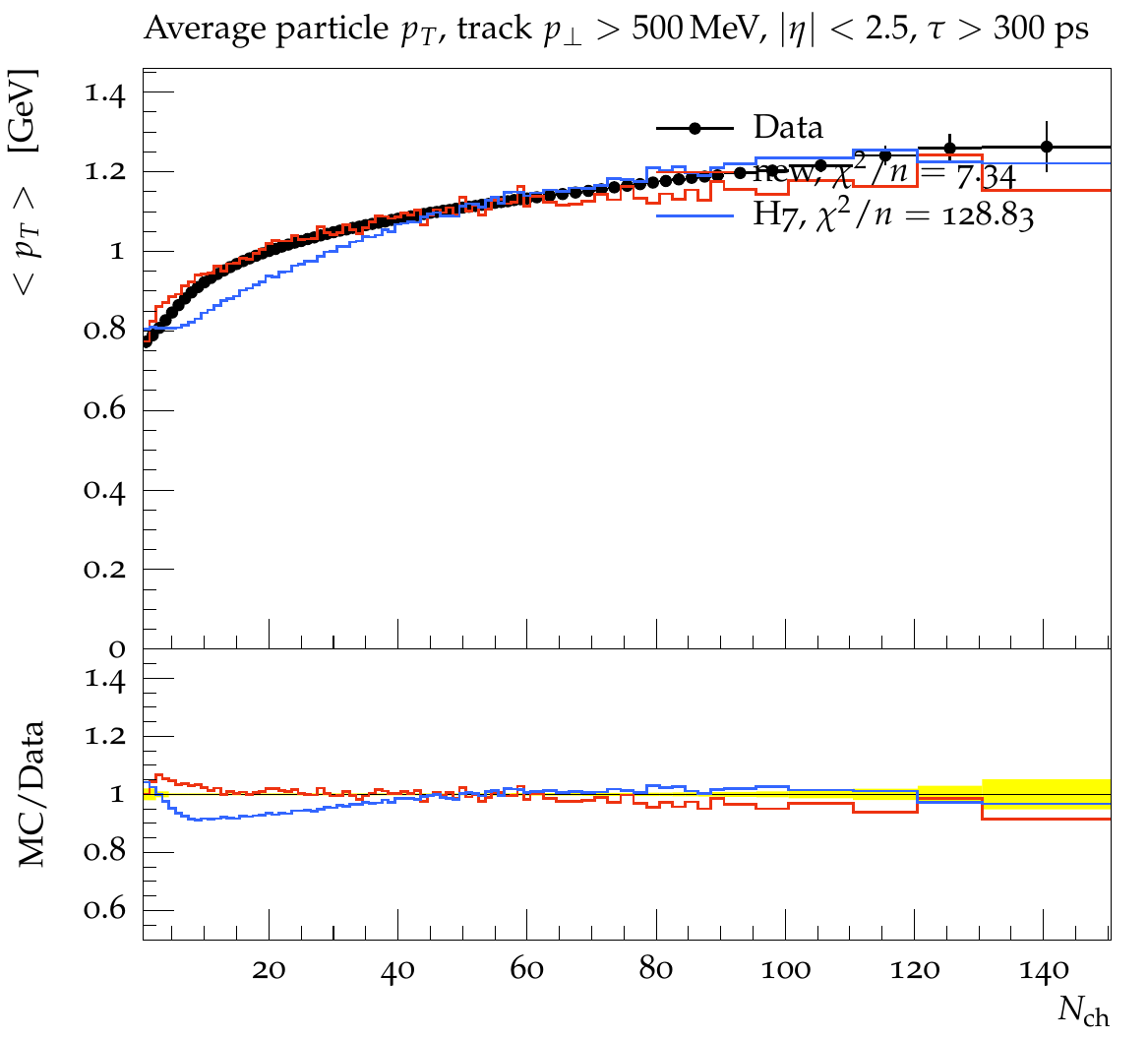}
	\caption{Most inclusive $\eta$ distribution for $p_{\perp}>500\, \rm{MeV}$ and average $p_{\perp}$ distribution for all particles with $p_{\perp}> 500 \, \rm{MeV}$ measured by ATLAS cite  at $\sqrt{s} = 13 \, \rm{TeV}$. The runs for the new model were simulated with the tuned set of parameters for $7\,\rm{TeV}$. H7.0 uses the old model for MPI \cite{Aad:2016mok}. }
	\label{13TeV}
\end{figure}

With the energy upgrade of the LHC to 13\,TeV in 2015, new sets of data are available. These may reveal insights into non-perturbative phenomena and also serve as important cross checks for the implementation of new models. To test the energy scaling of the model we run it with the same set of parameters as for $\sqrt{s} = 7\,\rm{TeV}$ and compare it to measurements from ATLAS\,\cite{Aad:2016mok}. The new model improves the description of the data significantly as shown in Fig.\,\ref{13TeV}. We note that with an identical set of parameters as for 7\,TeV we have good indication of a stable overall good energy scaling of our model.

\section{Conclusion}

In these proceedings we recap the implementation and explanation of the new model for soft interactions which was implemented in the Monte-Carlo event generator Herwig. It consists of a model for soft interactions and a model for diffraction. We showed that the new model was needed in order to resolve the so-called Bump-problem which was a clear artifact of the arbitrary colour connections between the gluons and the proton remnants in the old model for soft interactions and note severe improvements in all minimum-bias related observables considered. We also showed that the new model is able to give a reasonable description of underlying-event data and can be extrapolated to $13\, \rm{TeV}$.


\bibliography{bib}
\bibliographystyle{h-physrev}

\end{document}